\documentclass[twocolumn,floatfix,preprintnumbers,nofootinbib,superscriptaddress]{revtex4}

\usepackage{ulem}
\usepackage{bm}
\usepackage{times}
\usepackage{amssymb,amsbsy,amsmath,amsfonts}
\usepackage{graphicx}
\usepackage{float}
\usepackage{color}
\usepackage{morefloats}
\usepackage{rotating}
\usepackage{srcltx}
\usepackage{slashed}
\usepackage{subfigure}
\usepackage{multirow}
\usepackage{verbatim}
\usepackage{hyperref}
\usepackage{tabularx}

\usepackage[outdir=./]{epstopdf}

\usepackage{graphicx}
\usepackage{amsmath}
\usepackage{amsfonts}
\usepackage{amssymb}
\usepackage{color}
\usepackage{multirow}
\usepackage{mathrsfs}
\usepackage{gensymb}

\usepackage{booktabs}  
\usepackage{threeparttable}

\usepackage{subfigure}

\newcommand\be{\begin{eqnarray}}
\newcommand\ee{\end{eqnarray}}

\begin{document}

\title{Analysis of $B_s^0 \to X(3872) [\psi(2S)] \pi^+\pi^- (K^+ K^-)$ decays}

\author{Hao-Nan Wang}~\email{wanghaonan@impcas.ac.cn}
\affiliation{Institute of Modern Physics, Chinese Academy of Sciences, Lanzhou 730000, China}
\affiliation{School of Nuclear Sciences and Technology, University of Chinese Academy of Sciences, Beijing 101408, China}

\author{Li-Sheng Geng}~\email{lisheng.geng@buaa.edu.cn}
\affiliation{School of Physics, Beihang University, Beijing 102206, China}
\affiliation{Beijing Key Laboratory of Advanced Nuclear Materials and Physics, Beihang University, Beijing, 102206, China}
\affiliation{Peng Huanwu Collaborative Center for Research and Education, Beihang University, Beijing 100191, China}
\affiliation{Southern Center for Nuclear-Science Theory (SCNT), Institute of Modern Physics, Chinese Academy of Sciences, Huizhou 516000,  China}

\author{Gang Li}~\email{gli@qfnu.edu.cn}
\affiliation{College of Physics and Engineering, Qufu Normal University, Qufu 273165, China}

\author{Ju-Jun Xie}~\email{xiejujun@impcas.ac.cn}
\affiliation{Institute of Modern Physics, Chinese Academy of Sciences, Lanzhou 730000, China}
\affiliation{School of Nuclear Sciences and Technology, University of Chinese Academy of Sciences, Beijing 101408, China}
\affiliation{Southern Center for Nuclear-Science Theory (SCNT), Institute of Modern Physics, Chinese Academy of Sciences, Huizhou 516000, China}

\begin{abstract}

We have phenomenologically investigated the decays $B_s^0 \to  X(3872) \pi^+\pi^- (K^+ K^-)$ and $B_s^0  \to \psi(2S) \pi^+ \pi^- (K^+K^-)$. In our analysis, the scalar meson $f_0(980)$ is formed through the final state interactions of coupled channels $\pi \pi$ and $K\bar{K}$. Our findings indicate that the $\pi^+\pi^-$ invariant mass distribution of the $B_s^0 \to \psi(2S)\pi^+\pi^-$ decay can be accurately reproduced. Furthermore, we have explored the $\pi^+\pi^- (K^+ K^-)$ invariant mass distribution of the $B_s^0 \to X(3872) \pi^+\pi^- (K^+ K^-)$ decay, accounting for the different production mechanisms between $X(3872)$ and $\psi(2S)$, up to a global factor. It is found that the production rates for $X(3872)$ and $\psi(2S)$ are much different, which indicates that the structure of $X(3872)$ is more complicated than the $\psi(2S)$, which is a conventional $c\bar{c}$ state. Additionally, we have considered the contributions from $f_0(1500)$ to $\pi^+\pi^-$ and the $\phi$ meson to $K^+ K^-$ in our analysis. Utilizing the model parameters, we have calculated the branching fraction of $B_s^0  \to X(3872) K^+ K^-$, and anticipate that the findings of our study can be experimentally tested in the future.

\end{abstract}

\maketitle


\section{Introduction} \label{section:introduction}

The nonleptonic weak decays of bottom hadrons are widely acknowledged as a valuable means to elucidate the nature of certain enigmatic hadrons~\cite{Oset:2016lyh,Liang:2014tia,Liang:2014ama,Liang:2015twa,Xie:2018rqv,Liu:2020orv}, especially these decays with charmonia in the final states~\cite{LHCb:2015yax,Belle:2007hrb}. For example, it was found that the scalar meson $f_0(500)$ has a relatively bigger signal than $f_0(980)$ in the decay of $\bar{B}^0$ into $J/\psi\pi^+\pi^-$~\cite{LHCb:2013dkk}. While the decay of $B_s^0 \to J/\psi \pi^+\pi^-$ was measured by the LHCb collaboration~\cite{LHCb:2011blx}, and a pronounced peak was found for the scalar meson $f_0(980)$ in the $\pi^+\pi^-$ invariant mass distributions. However, there was no appreciable signal for the scalar meson $f_0(500)$~\cite{LHCb:2011blx}. This counter-intuitive result attracted experimental and theoretical attention. New measurements about the $B$ and $B_s$ decays have been performed by the Belle Collaboration~\cite{Belle:2011phz}, CDF Collaboration~\cite{CDF:2011kjt}, D0 Collaboration~\cite{D0:2011zbi}, and LHCb Collaboration~\cite{LHCb:2012ae,LHCb:2014ooi}. 

The $B^0_s \to J/\psi \pi^+\pi^-$ decay was studied in Ref.~\cite{Liang:2014tia} based on the final state interaction of pseudo-scalar meson-pseudo-scalar meson provided by the chiral unitary approach, where the scalar mesons $f_0(500)$ and $f_0(980)$ were dynamically generated. The theoretical results are in agreement with the experimental data~\cite{LHCb:2011blx}. The approach of Ref.~\cite{Liang:2014tia} was successfully extended to study other weak decays of $B^0_s$ and $B$ mesons~\cite{Xie:2014gla,Liang:2015qva,Dai:2015bcc,Albaladejo:2016hae,Molina:2016pbg,Bayar:2014qha} (see also Ref.~\cite{Oset:2016lyh} for an extensive review). The $B_s^0 \to \psi(2S)\pi^+\pi^-$ decay was firstly measured by the LHCb collaboration~\cite{LHCb:2013mrs} and the $f_0(980)$ meson played an important role in the $\pi^+\pi^-$ invariant mass distributions. Recently, the $B^0_s \to X(3872) \pi^+ \pi^-$ decay was also firstly observed by the LHCb collaboration~\cite{LHCb:2023reb}, where a large contribution from $B^0_s \to X(3872) [f_0(980) \to \pi^+\pi^-]$ was found. Determining the $f_0(980)$ nature in the $B^0_s$ decays is possible. Indeed, it is interesting to investigate $f_0(980)$ in $B_s^0 \to X(3872)\pi^+\pi^-$, since it is the analogous decay compared with the decay of $B_s^0$ into $\psi(2S)\pi^+\pi^-$ within the assumption that $X(3872)$ can be generated by the hadronization of $c\bar{c}$ which is used to produce $\psi(2S)$ in the former case. The $B^0_s \to X(3872)\pi^+\pi^-$ decay is also a useful platform to explore the exotic feature of $X(3872)$~\cite{Chen:2022asf,Dong:2021bvy,Guo:2019twa,Ali:2017jda,Olsen:2017bmm,Esposito:2016noz,Chen:2016qju,Lebed:2016hpi,Wang:2022xga,Liang:2024sbw}. Even if it was discovered about two decades ago~\cite{Belle:2003nnu,CDF:2003cab,D0:2004zmu,CDF:2006ocq,BaBar:2008qzi,CDF:2009nxk,LHCb:2011zzp,CMS:2013fpt,LHCb:2013kgk,LHCb:2015jfc}, its nature is still unclear. For instance, molecular perspective is one common explanation for $X(3872)$ rather than a pure chamonium. As discussed in Ref.~\cite{Liang:2024sbw}, one has investigated the decays of $B$ meson into $X(3872)$ with a pseudoscalar or vector meson based on the molecular perspective of $X(3872)$ from the interaction of $D\bar{D}^*+c.c.$(charge conjugate). Following the analysis about the $B_s^0 \to \psi(2S) \pi^+\pi^-$ decay, we will also study the $B_s^0 \to X(3872) \pi^+\pi^-$ decay.

It is natural to study the role of $f_0(980)$ in the $K^+K^-$ invariant mass distribution of $B_s^0 \to \psi(2S)K^+K^-$ and $B_s^0 \to X(3872)K^+K^-$ decays using the chiral unitary approach, since $f_0(980)$ has strong coupling to the $K\bar{K}$ channel~\cite{Oller:1997ti,Guerrero:1998ei}. Note that, within the chiral unitary approach~\cite{Nieves:1999bx,Pelaez:2006nj,Kaiser:1995eg,Oller:2000ma}, the production of $f_0(980)$ and $f_0(500)$ mesons in $B^0$ and $B_s^0$ into $J/\psi$ and a $\pi^+\pi$ or $K^+K^-$ pair were investigated in Refs.~\cite{Bayar:2014qha,Liang:2014tia,Xie:2018rqv}. To understand the new experimental data collected by the LHCb collaboration~\cite{LHCb:2023reb} and study the nature of $X(3872)$ and the scalar meson $f_0(980)$, in this work, we perform a coherent analysis of the $B^0_s \to X(3872) \pi^+ \pi^- (K^+ K^-)$ and $B^0_s \to \psi(2S) \pi^+ \pi^- (K^+ K^-)$ decays. In addition to the $f_0(980)$, we also consider the contribution from the scalar meson $f_0(1500)$, since its signal is clearly seen in the invariant $\pi^+\pi^-$ mass distributions~\cite{LHCb:2023reb,Ochs:2013vxa,Ochs:2013gi}.

This paper is organized as follows. In Sec.~\ref{sec:formalism}, we present the theoretical formalism for the production of the scalar meson $f_0(980)$ in the $B_s^0$ decays into $\psi(2S)$ or $X(3872)$ and $\pi^+\pi^-$ or $K^+K^-$, together with a discussion about the scalar meson $f_0(1500)$ in the corresponding decays, while the contribution of the $\phi$ meson in the $B_s^0 \to X(3872)K^+K^-$ decay is also shown. In Sec.~\ref{sec:results}, we show our theoretical numerical results and discussions, followed by a summary in the last section.


\section{Theoretical formalism} \label{sec:formalism}

\subsection{The $B_s^0 \to \psi(2S) [f_0(980), f_0(1500) \to \pi^+\pi^-]$ decay} \label{subsec:2.1}

\begin{figure}[htbp]
\centering
\subfigure{\includegraphics[scale=0.45]{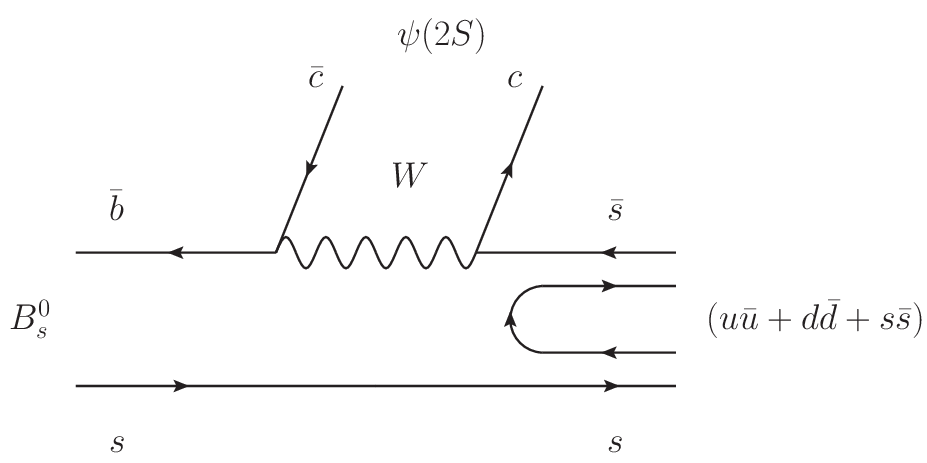}}
\caption{Diagram for the decay of $B_s^0$ into $\psi(2S)$ (formed by the $c\bar{c}$ pair) and a primary $s\bar{s}$ pair, which hadronizes with an extra $(u \bar{u} + d \bar{d} + s\bar{s})$ pair from the vacuum.} \label{fig:980pipi}
\end{figure}

The leading contributions to the decays of $B^0_s$ into $\psi(2S)$ plus a scalar meson is the Cabibbo favored $\bar{b} \to c \bar{c} \bar{s}$ process, therefore, the decay diagram of $B_s^0 \to \psi(2S) [f_0(980) \to \pi^+\pi^-]$, at the quark level, is shown in Fig.~\ref{fig:980pipi}, which can be separated into two steps. The first step, namely the Cabbibo favored process, consists of the $\bar{b}$ decaying into a $\bar{c}$ quark and a $W^+$ boson followed by its decay into a $c$ quark and a $\bar{s}$ quark. Then, in addition to the hadronization of $c\bar{c}$ to produce $\psi(2S)$, we need another $q\bar{q}$ ($\equiv u \bar{u} + d\bar{d} + s\bar{s}$) pair to generate the $\pi^+\pi^-$ in the final states from $s\bar{s}$.

Following Refs.~\cite{Liang:2014tia,Xie:2018rqv}, the hadronization of $s\bar{s}$, in terms of pseudoscalar mesons, can be written as

\be 
s\bar{s}(u\bar{u}+d\bar{d}+s\bar{s}) \to  K^+K^- + K^0\bar{K}^0 + \frac{2}{3}\eta\eta .
\label{eq:hadronization}
\ee 
After the pseudoscalar meson-pseudoscalar meson pair is produced, final-state interactions between the mesons occur, where the $\pi^+\pi^-$ pair can be obtained in the final states. The scalar meson $f_0(980)$ is dynamically generated from the $s$-wave interaction of the pseudoscalar meson-pseudoscalar meson in coupled channels~\cite{Oller:1998hw,Kaiser:1998fi,Locher:1997gr}. Hence, the decay amplitude for $B_s^0 \to \psi(2S) [f_0(980) \to \pi^+\pi^-]$ can be written as~\cite{Liang:2014tia},
\be 
    && \mathcal{M}_{B_s^0 \to \psi(2S)\pi^+\pi^-}^{f_0(980)} = g_1 
 \mathcal{M}_a =  \frac{g_1 V_{cs} \boldsymbol{p}_{\psi(2S)} \cos{\theta}}{m_{B_s^0}} \times  \notag\\
 && \Big( G_{K^+K^-}t_{K^+K^- \to \pi^+\pi^-}  
    + G_{K^0\bar{K}^0}t_{K^0\bar{K}^0 \to \pi^+\pi^-} \notag\\ &&  + \frac{2}{3}\frac{1}{2}G_{\eta\eta}t_{\eta\eta \to \pi^+\pi^-} \Big),
\label{eq:980pipi}
\ee 
where $\boldsymbol{p}_{\psi(2S)}$ is the three momentum of $\psi(2S)$ in the center-mass system of $B_s^0$ and $\theta$ is an integration variable of final-state phase space. Note that for the $B_s^0 \to \psi(2S) [f_0(980) \to \pi^+\pi^-]$ decay, we shall need a $p$-wave interaction to match angular momentum conservation. We introduce a parameter $g_1$ to contain all dynamical factors, which is assumed to be real and positive in this work. The $V_{cs}$ is one  matrix element of the Cabbibo-Kobayashi-Maskawa matrix which is related to the Cabbibo angle~\cite{Bayar:2014qha}: $V_{cs} = \cos{\theta_c} = 0.97427$.

In Eq.~(\ref{eq:980pipi}), $G_i$ is the loop function of two meson propagators
\be 
    G_i(s) = {\rm i}\int\frac{{\rm d}^4q}{(2\pi)^4}\frac{1}{(P-q)^2-m_1^2+{\rm i}\varepsilon}\frac{1}{q^2 - m_2^2 + {\rm i}\varepsilon},
\label{eq:loopfunction}
\ee 
where ``$i$" represents the $i$th-channel, $m_1$, $m_2$, and $q$ are the masses and four-momentum of one meson in this channel, respectively. $P$ is the total momentum in this system, satisfying $s = P^2$. The three-momentum integral is carried out by precisely integrating the $q^0$ variable and applying a cutoff $\Lambda$ of the order of 1 GeV, which is impacted by the number of channels.  The element of the scattering matrix, $t_{ij}$, for the transition of channel $i$ to $j$, is given by $t = (1 - VG)^{-1}V$. Now numbering the channels as 1 for $\pi^+\pi^-$, 2 for $\pi^0\pi^0$, 3 for $K^+K^-$, 4 for $K^0\bar{K}^0$, and 5 for $\eta\eta$, the $V$ matrix can be used in the same form as~\cite{Liang:2014tia}. It is worth noting whether or not considering the $\eta\eta$ channel does not affect the results much, as long as a reasonable cutoff $\Lambda$ is used. See more details in Refs.~\cite{Liang:2014tia,Oller:1997ti,Kaiser:1998fi}. We don't consider the $\eta\eta$ channel in this work and take $\Lambda = 903$ MeV. The loop function $G$ and two-body scattering amplitude $t$ depend on the invariant mass $M_{\pi \pi}$ of the $\pi^+ \pi^-$ system.

\begin{figure}[htbp]
	\centering
	\subfigure{\includegraphics[scale=0.7]{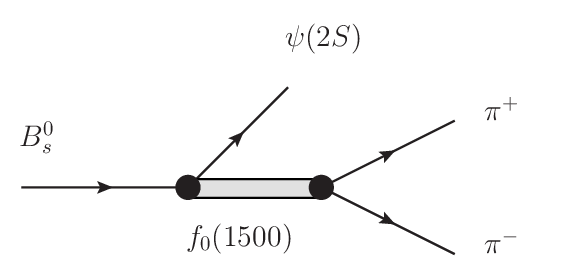}}
	\caption{Diagram for the decay of $B_s^0$ into $\psi(2S)$ and $\pi^+\pi^-$ through the resonance $f_0(1500)$.} \label{fig:1500}
\end{figure}

In addition to the scalar meson $f_0(980)$, we consider the scalar meson, namely $f_0(1500)$ as shown in Fig.~\ref{fig:1500}. It is treated in the amplitude as a Breit-Wigner (BW) propagator.
\be 
    && \mathcal{M}_{B_s^0 \to \psi(2S)\pi^+\pi^-}^{f_0(1500)} = g_2\mathcal{M}_b \notag\\
    &=& \frac{{\rm i}g_2 m_{f_0(1500)}\Gamma_{f_0(1500)} \boldsymbol{p}_{\psi(2S)}\cos{\theta}}{m_{B_s^0} \big( M^2_{\pi \pi} - m_{f_0(1500)}^2 + {\rm i}m_{f_0(1500)}\Gamma_{f_0(1500)} \big)},   \label{eq:1500}
\ee 
where $m_{f_0(1500)}$ and $\Gamma_{f_0(1500)}$ are the mass and width of $f_0(1500)$. Here, $g_2$ is a free parameter, and we consider it real and positive. Furthermore, ongoing debates exist about the nature of $f_0(1500)$, and its mass and width are not well determined~\cite{Workman:2022ynf}. Hence, $m_{f_0(1500)}$ and $\Gamma_{f_0(1500)}$ will be fitted to the experimental data.

Then, the total decay amplitude for $B^0_s \to \psi(2S)\pi^+\pi^-$ is written as,
\be 
    \mathcal{M}_{B^0_s \to \psi(2S)\pi^+\pi^-} = g_1\mathcal{M}_a + g_2\mathcal{M}_b e^{{\rm i}\varphi},
    \label{eq:psipipi}
\ee
where $\varphi$ is the relative phase between ${\cal M}_a$ and ${\cal M}_b$, and it is a free parameter. In fact, as discussed in Ref.~\cite{LHCb:2023reb}, there are indeed contributions from the interference between $f_0(980)$ and $f_0(1500)$ to the $\pi^+\pi^-$ invariant mass spectrum of the $B_s^0 \to \psi(2S)\pi^+\pi^-$ decay.

\subsection{The mechanism of $B_s^0 \to X(3872)\pi^+\pi^- (K^+ K^-)$} \label{subsec:2.2}

In contrast with the charmonium state $\psi(2S)$, the production of $X(3872)$ in the decay of $B_s^0 \to X(3872)\pi^+\pi^-$ may have a more involved mechanism because of the exotic nature of the $X(3872)$ state. Therefore, we should involve a different parameter $g'_1$ [see  Eq.~(\ref{eq:980pipi})] for the $B_s^0 \to X(3872)\pi^+\pi^-$ decay:~\footnote{Note that the $\eta\eta$ channel is also neglected.}
\be 
    && \mathcal{M}_{B_s^0 \to X(3872)\pi^+\pi^-} = \frac{g_1'V_{cs}\boldsymbol{p}_{X(3872)}\cos{\theta}}{m_{B_s^0}} \times  \notag\\
   &&   \Big( G_{K^+K^-}t_{K^+K^- \to \pi^+\pi^-} + G_{K^0\bar{K}^0}t_{K^0\bar{K}^0 \to \pi^+\pi^-}  \Big).
\label{eq:3872pipi}
\ee 
In other words, the mechanism for the production of $X(3872)$ is the same as that shown in Fig.~\ref{fig:980pipi} if we only consider the short-range contribution to the hadronization of $c\bar{c}$.

On the other hand, the contribution of $f_0(1500)\to \pi^+\pi^-$ in $B_s^0 \to X(3872)\pi^+\pi^-$ is different from that in the decay $B_s^0 \to \psi(2S)\pi^+\pi^-$. Referring to the masses of relevant particles in the Review of Particle Physics (RPP)~\cite{Workman:2022ynf}, the phase space is tiny for the $B_s^0 \to X(3872)\pi^+\pi^-$ decay. The upper limit of the invariant mass of $M_{\pi\pi}$ is barely bigger than the mass of $f_0(1500)$, which means that the peak of $f_0(1500)$ in the $\pi^+\pi^-$ invariant mass distribution is seriously suppressed. Even if there is some contribution from $f_0(1500)$, it can be omitted in our mechanism. Thus, Eq.~(\ref{eq:3872pipi}) is essentially the complete amplitude of the $B_s^0 \to X(3872)\pi^+\pi^-$ decay.

\begin{figure}[htbp]
\centering
\subfigure[]{\includegraphics[scale=0.5]{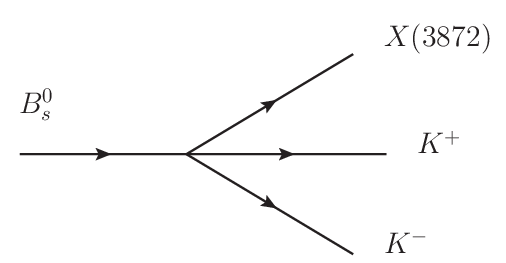}} 
\subfigure[]{\includegraphics[scale=0.5]{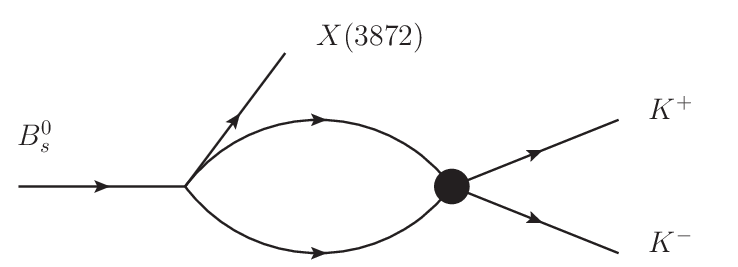}}
\caption{Diagram for the decay of $B_s^0 \to X(3872) K^+K^-$ where $K^+K^-$ is produced in $s$-wave. (a) is the tree diagram, and (b) is the rescattering.} \label{fig:3}
\end{figure}

The $K^+K^-$ pair can not only be directly produced by the hadronization of $s\bar{s}$ with $u\bar{u}$ from the vacuum in Fig.~\ref{fig:980pipi},  but also be dynamically produced by the final-state interaction of $K \bar{K}$ in $s$-wave. According to the diagrams shown in Fig.~\ref{fig:3}, the decay amplitude of $B_s^0 \to X(3872)f_0(980) \to X(3872) K^+ K^-$ is given by
\be 
    && \mathcal{M}_{B_s^0 \to X(3872) K^+K^-} =\frac{g_1'V_{cs}\boldsymbol{p}_{X(3872)}\cos{\theta}}{m_{B_s^0}} \Big( 1 + \notag\\
    && G_{K^+K^-}t_{K^+K^- \to K^+ K^-} + G_{K^0\bar{K}^0}t_{K^0\bar{K}^0 \to K^+ K^-}  \Big),
\label{eq:3872KK}
\ee 
where we have used the same coupling constant $g_1'$ as in Eq.~(\ref{eq:3872pipi}) because of the similar mechanism and the same final state $X(3872)$.

On the other hand, we also consider the contribution of the $\phi$ meson to the $B_s^0 \to X(3872) K^+ K^-$ decay. In this case, the $K^+K^-$ is produced in $p$-wave. The decay amplitude is written as,
\be 
    && \mathcal{M}^{\phi}_{B_s^0 \to X(3872) K^+ K^-} = g_{BX\phi}g_{\phi KK}\varepsilon^{\mu\nu\rho\sigma} \notag\\ 
    && \times \epsilon_{\mu}^*(\boldsymbol{p}_X)p_{X\nu}q_{\sigma} \frac{{\rm i}(p_{K^+} - p_{K^-})_{\rho}}{q^2 - m_{\phi}^2 + {\rm i}m_{\phi}\Gamma_{\phi}} ,
\label{eq:phiKK}
\ee 
where $\epsilon_{\mu}^*(\boldsymbol{p}_X)$ and $p_X$ are the polarization and four momentum of $X(3872)$. And $\epsilon_{\nu}^*(\boldsymbol{q})$, $q$, $m_{\phi}$ and $\Gamma_{\phi}$ are the polarization, four-momentum, mass, and width of the $\phi$ meson. Besides, $g_{BX\phi}$ and $g_{\phi KK}$ are the coupling parameters of the vertexes of $B_s^0X(3872)\phi$ and $\phi KK$. With the branching fractions of $\mathcal{B}[{B_s^0 \to X(3872)\phi}] = (1.1\pm0.4)\times10^{-4}$ and $\mathcal{B}[{\phi \to K^+K^-}] = (49.1\pm0.5)\%$ from RPP~\cite{Workman:2022ynf}, one can obtain that $g_{BX\phi}^2 = (7.3\pm2.7)\times 10^{-22}\text{MeV}^{-2}$ and $g_{\phi KK}^2 = (20.0 \pm 0.2)$.

\subsection{The mass distribution and partial decay width of $B_s^0 \to X(3872) [\psi(2S)] \pi^+\pi^- (K^+ K^-)$} \label{subsec:2.4}

With these decay amplitudes obtained above, the $\pi^+ \pi^-$ and $K^+K^-$ invariant mass distributions of $B_s^0 \to X(3872) [\psi(2S)] \pi^+\pi^- (K^+ K^-)$ decay can be easily obtained as follows
\be 
    \frac{{\rm d}\Gamma}{{\rm d} M_{\rm inv}} &=& \frac{1}{512 \pi^5 m_{B_s^0}^2} \int {\rm d}\Omega \Omega^* |\boldsymbol{p}||\boldsymbol{p}^*|  |\mathcal{M}|^2 ,
\label{eq:massdistribution}
\ee 
where $(\boldsymbol{p}, \Omega)$ is the three momentum of $X(3872)$ or $\psi(2S)$ in the rest frame of $B^0_s$, while $(\boldsymbol{p}^*,\Omega^*)$ is the three momentum of one $\pi$ ($K$) in the final $\pi^+ \pi^-$ ($K^+K^-$) center of mass frame with invariant mass $M_{\rm inv}$. 


\section{Numerical results and discussions}\label{sec:results}

We calculate the invariant $\pi^+\pi^-$ mass distributions of the process $B^0_s \to \psi(2S) \pi^+\pi^-$ with the above theoretical formalism. There are five free parameters to be obtained by fitting to the
experimental data: $g_1$ in Eq.~(\ref{eq:980pipi}), $g_2$, $m_{f_0(1500)}$, and $\Gamma_{f_0(1500)}$ in Eq.~(\ref{eq:1500}), and $\varphi$ in Eq.~(\ref{eq:psipipi}). Since our numerical results are ${\rm d}\Gamma/{\rm d}M_{\rm inv}$, and the experimental data are events as a function of the $\pi^+ \pi^-$ invariant mass, there is a global factor $C$ between our theoretical calculations and the experimental data. On the other hand, in the fitting to the experimental data, we use the following form:
\begin{eqnarray} 
  &&  {\rm data} = C \frac{{\rm d}\Gamma}{{\rm d}M_{\rm inv}} 
       = \frac{Cg_2^2}{512 \pi^5 m_{B_s^0}^2}  \int {\rm d}\Omega \Omega^* |\boldsymbol{p}||\boldsymbol{p}^*| \times \notag\\
       && \Big[ (\frac{g_1}{g_2})^2|\mathcal{M}_a|^2 + |\mathcal{M}_b|^2 + \frac{2g_1}{g_2}  {\rm Re} \Big(\mathcal{M}_a^*\mathcal{M}_b e^{{\rm i}\varphi} \Big) \Big].
    \label{eq:fitmodified}
\end{eqnarray} 
Thus, we take $Cg_2^2$, $g_1/g_2$, $m_{f_0(1500)}$, $\Gamma_{f_0(1500)}$, and $\varphi$ as free parameters. It should be noted that the global factor $C$ can normalize the theoretical results to match the experimental mass distribution. And more importantly, the factor $C$ is the same for the two processes $B_s^0$ into $\psi(2S)\pi^+\pi^-$ and $X(3872)\pi^+\pi^-$. In this way, the fitted parameters are listed in Table~\ref{table:fitparameter}. The obtained $\chi^2/{\rm d.o.f}$ is 1.4, which is reasonably small.

\begin{table}[htbp]
\renewcommand\arraystretch{1.5}
\centering
		\caption{The fitted parameters in this work.}  
		\begin{tabular}{cc}   
		\toprule[1.5pt]
			~~~~~~~~Parameters~~~~  & ~~~~Fitting Results~~~~~~~~  \\
			 \hline
			$Cg_2^2$  & $(2.77 \pm 0.35)\times 10^{8}$  \\
                $g_1/g_2$  & $0.68 \pm 0.04$  \\
			$\varphi(\degree)$  & $-85.11 \pm 8.65$  \\
			$m_{f_0(1500)}$ (MeV)  & $1450.0 \pm 6.8$  \\
                $\Gamma_{f_0(1500)}$ (MeV)  & $164.4 \pm 22.4$  \\
              \hline
			$\chi^2/{\rm d.o.f}$  & 1.4  \\
	\toprule[1.5pt]
		\end{tabular}    
\label{table:fitparameter}
\end{table}

\begin{figure}[htbp]
\centering
\includegraphics[scale=0.3]{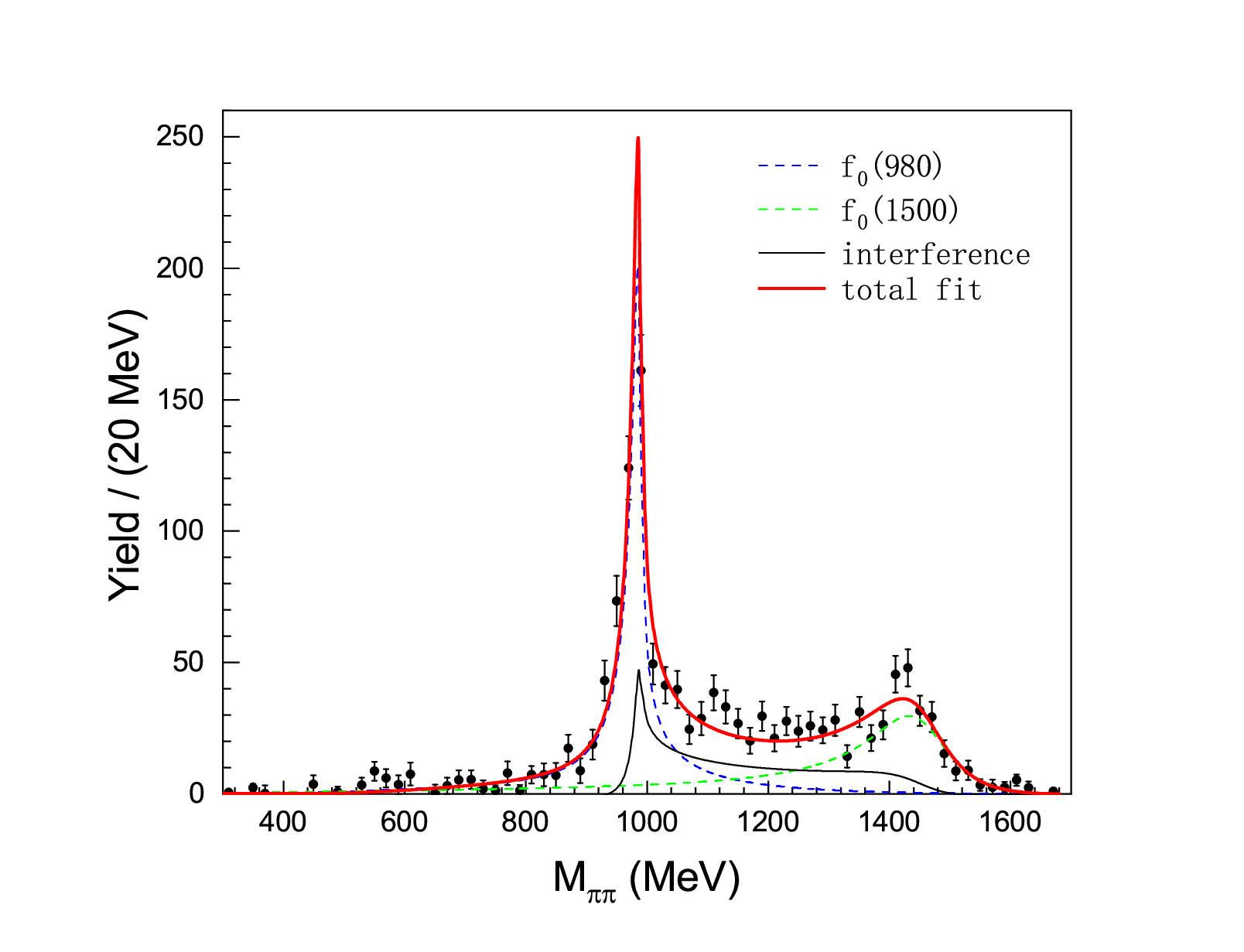}
\caption{Invariant mass distribution of $\pi^+\pi^-$ for the $B_s^0 \to \psi(2S)\pi^+\pi^-$ decay, compared with the experimental data taken from Ref.~\cite{LHCb:2023reb}. The blue-dashed, green-dashed, and black-solid curves are the contributions from the $f_0(980)$, $f_0(1500)$, and their interference, respectively. The red-solid line is their total contribution.} \label{fig:fitline}
\end{figure}

The fitted results of the $\pi^+ \pi^-$ invariant mass distributions of $B_s^0 \to \psi(2S)\pi^+\pi^-$ are shown in Fig.~\ref{fig:fitline}. One can see that thanks to the contributions from $f_0(980)$ and $f_0(1500)$, the experimental data can be well reproduced. In the calculations, the scalar meson $f_0(980)$ is produced in the final-state interaction of $K\bar{K}$ and $\pi \pi$ in coupled channels. The first higher peak can be described by only the $f_0(980)$ state. In contrast, the second small peak and the long tail between the two peaks can be reproduced by the $f_0(1500)$ and the interference between $f_0(980)$ and $f_0(1500)$. It is worth mentioning that the mass and width of $f_0(1500)$ state are mainly determined by the second peak, and the fitted results are different from the values quoted in the RPP~\cite{Workman:2022ynf}.

With the fitted parameters and the branching ratio of $ \mathcal{B}[B_s^0 \to \psi(2S)\pi^+\pi^-] = (6.9 \pm 1.2) \times 10^{-5}$~\cite{Workman:2022ynf}, we can extract the global factor $C$, which is $C = (8.28 \pm 1.44)\times 10^{17}$. Then, we can also get $g_2 = (1.83 \pm 0.20)\times 10^{-5}$ and $g_1 = (1.24 \pm 0.15)\times 10^{-5}$. If we take the same coupling constant $g_1$ for the $B_s^0 \to J/\psi \pi^+\pi^-$ decay, we obtain $\Gamma[B_s^0 \to J/\psi f_0(980) \to J/\psi \pi^+\pi^-] = (3.9 \pm 1.0) \times 10^{-14}$ MeV, which is in agreement with the value of $(5.4 \pm 0.6)\times 10^{-14}$ MeV quoted in the RPP~\cite{Workman:2022ynf}. This indicates that the coupling constants for producing charmonium states in the $B^0_s$ decays are universal.

\begin{figure}[htbp]
\centering
\includegraphics[scale=0.3]{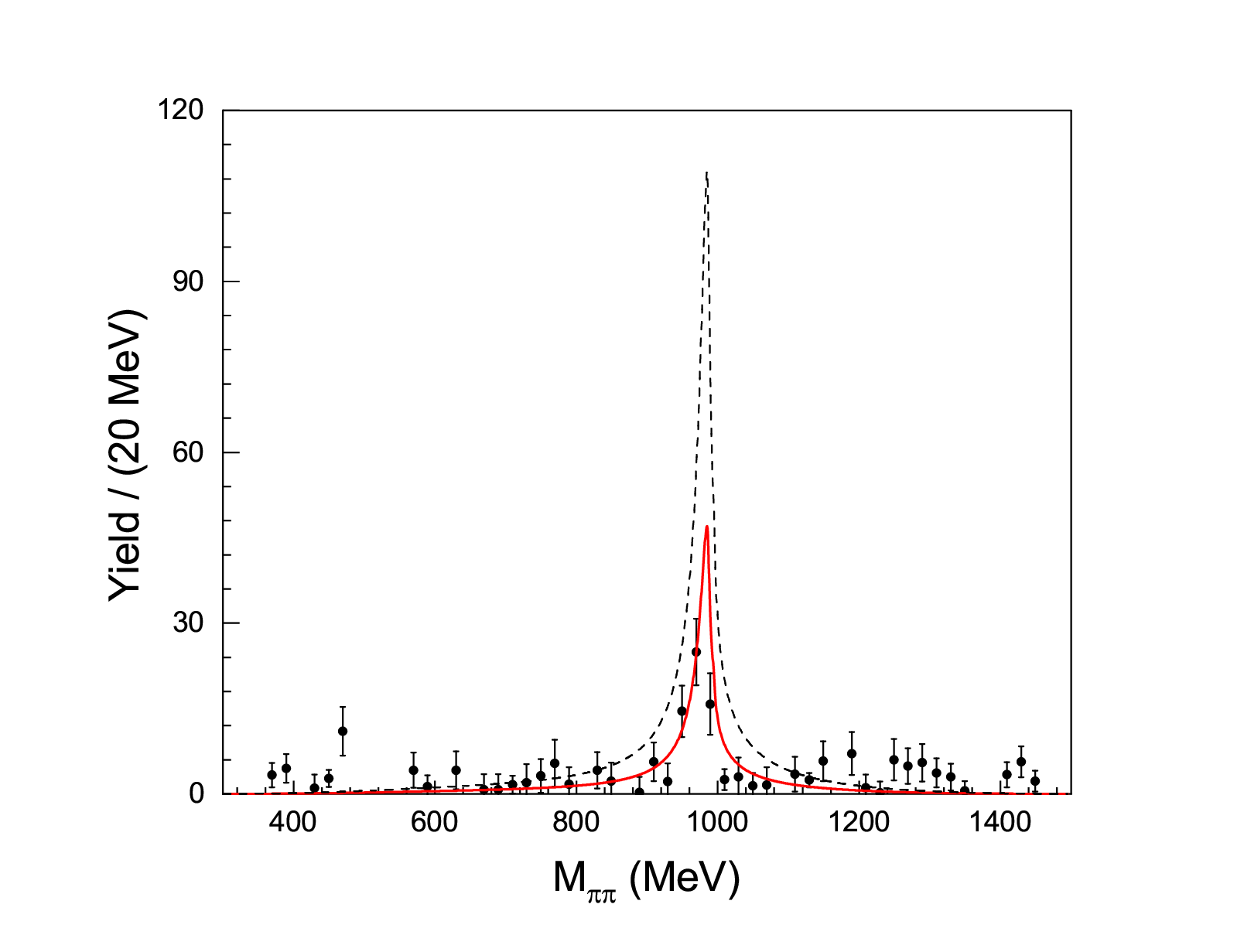}
\caption{Invariant mass distribution of $\pi^+\pi^-$ for the $B_s^0 \to X(3872)\pi^+\pi^-$ decay, compared with the experimental data~\cite{LHCb:2023reb}. The red-solid and black-dashed curves are obtained with different values for the production parameter $g_1'$.} \label{fig:3872pipi}
\end{figure}

Next, we turn to the $B_s^0 \to X(3872)\pi^+\pi^-$ decay. We firstly set $g_1' = g_1$. The resulting invariant mass $M_{\pi\pi}$ distribution of $B_s^0 \to X(3872)\pi^+\pi^-$ is shown as the black-dashed curve in Fig.~\ref{fig:3872pipi}. In this case, the obtained peak of $f_0(980)$ is too high compared with the available experimental data around $980$ MeV. This indicates that the coupling of $g_1'$ should differ from that of $g_1$. In another words, the production mechanism of $X(3872)$ and $\psi(2S)$ in the $B^0_s \to X(3872) [\psi(2S)] \pi^+ \pi^-$ decays are different. Indeed, the contributions from the long-distance $\bar{D}D^*$ scattering to the $X(3872)$ production in the $B^0_s$ decays are important~\cite{Wang:2022xga,Liang:2024sbw}.

To get a good description of the experimental data on the $B^0_s \to X(3872) \pi^+ \pi^-$ decay, we modify the value of $g_1'$ and enable the theoretical results to pass through the highest experimental point around $M_{\pi\pi} = 980$ MeV. We get $g_1' = 0.69 g_1$, and the corresponding results are shown in Fig.~\ref{fig:3872pipi} by the red curve. To explore more details about the difference between the production of $X(3872)$ and the charmonium states in the $B^0_s$ decays, we compare the modulus squared of their decay amplitudes, where the effect of the phase space is removed. For this purpose, we write,
\be 
     |\mathcal{M}_{B_s^0 \to R f_0(980)}|^2 = \Gamma_{B_s^0 \to R f_0(980)}/|\boldsymbol{p}_{R}|,
\label{eq:twobodyam}
\ee 
where $R$ represents the $X(3872)$, $\psi(2S)$, or $J/\psi$, respectively. For the partial decay width of $\Gamma_{B_s^0 \to R f_0(980)}$, we calculate them with the spectral function for the $\pi^+ \pi^-$ distribution as follows~\cite{Bayar:2014qha,Liang:2014ama},
\be 
    \Gamma_{B_s^0 \to R f_0(980)} =  \frac{\int^{M^{\rm max}_{\pi\pi}}_{M^{\rm min}_{\pi \pi}} \frac{ {\rm d}\Gamma_{B_s^0 \to R \pi^+\pi^-} } {dM_{\pi \pi}} /S(M^2_{\pi\pi}) dM_{\pi \pi} }{  \int^{M^{\rm max}_{\pi\pi}}_{M^{\rm min}_{\pi \pi}}  dM_{\pi \pi}},
\label{eq:spectral}
\ee 
with the special function $S(M^2_{\pi\pi})$~\footnote{Here, we use a Breit-Wigner form for the $f_0(980)$, and it will not change our main conclusion if we worked in the dynamically generated picture.},
\be 
S(M^2_{\pi \pi})=  - {\rm Im}\frac{2m_{f_0(980)}/\pi}{M_{\pi\pi}^2 - m_{f_0(980)}^2 + {\rm i}m_{f_0(980)}\Gamma_{f_0(980)}} ,
\ee
where we take $m_{f_0(980)} = 985$ MeV and $\Gamma_{f_0(980)} = 100$ MeV as quoted in the RPP~\cite{Workman:2022ynf}, while $M^{\rm max}_{\pi\pi} = 1035~{\rm MeV}$ and $M^{\rm min}_{\pi \pi} = 935~{\rm MeV}$. 

On the other hand, we can also evaluate the modulus squared of decay amplitudes for $B^0_s \to R \phi (\eta, \eta')$ by replacing $f_0(980)$ with $\phi$, $\eta$ or $\eta'$, which contain an $s\bar{s}$ component. We define:
\be 
R_1 &=& \frac{|\mathcal{M}_{B_s^0 \to X(3872) f_0(980) [\phi, \eta, \eta']}|^2}{|\mathcal{M}_{B_s^0 \to J/\psi f_0(980) [\phi, \eta, \eta']}|^2}, \label{eq:ratio1}\\
R_2 &=& \frac{|\mathcal{M}_{B_s^0 \to \psi(2S) f_0(980) [\phi, \eta, \eta']}|^2}{|\mathcal{M}_{B_s^0 \to J/\psi f_0(980) [\phi, \eta, \eta']}|^2}.
\label{eq:ratio2}
\ee 
These obtained numerical results for $R_1$ and $R_2$ are listed in Table~\ref{table:ratio2}. In the calculations, we take the two-body decay branching fractions from the RPP~\cite{Workman:2022ynf} except $\mathcal{B}[B_s^0 \to X(3872)\eta]$ and $\mathcal{B}[B_s^0 \to X(3872) \eta']$. While for the decays of $B_s^0 \to X(3872)\eta$ and $B_s^0 \to X(3872) \eta'$, there are still no experimental measurements, and thus we rely on the results obtained in Ref.~\cite{Wang:2022xga} with the subtraction parameter $\alpha = -1.91$ (see more details in that reference). These values are listed in Table~\ref{table:Br}. On the other hand, Ref.~\cite{Liang:2024sbw} also gives the results of $\mathcal{B}[B_s^0 \to X(3872)\eta]$ and $\mathcal{B}[B_s^0 \to X(3872) \eta']$ based on the molecular picture of $X(3872)$.

\begin{table}[htbp]
\renewcommand\arraystretch{1.5}
\centering
		\caption{Ratios of the two-body decays of $B_s^0 \to X(3872)[\psi(2S)]f_0(980)[\phi, \eta, \eta']$ to the $B_s^0 \to J/\psi f_0(980)[\phi, \eta, \eta']$.}  
		\begin{tabular}{ccc}   
		\toprule[1.5pt]
              ~~~~~~~~~~~~~~~~&~~~~~~ $R_1$ ~~~~~~&~~~~~~ $R_2$ ~~~~~~~~~~~~ \\ 
			 \hline
			$f_0(980)$  & $0.16 \pm 0.09$ & $0.51 \pm 0.28$ \\
                $\phi$  & $0.18 \pm 0.07$ & $0.71 \pm 0.06$ \\
			$\eta$  & $0.05 \pm 0.03$& $1.07 \pm 0.35$ \\
			$\eta'$ & $0.08 \pm 0.04$ & $0.54 \pm 0.16$\\
	\toprule[1.5pt]
		\end{tabular}    
\label{table:ratio2}
\end{table}

\begin{table}[htbp]
\renewcommand\arraystretch{1.5}
\centering
		\caption{Branching fractions of $B_s^0$ decaying into $X(3872)[\psi(2S), J/\psi]$ and $\phi[\eta, \eta']$.}  
		\begin{tabular}{cc}   
		\toprule[1.5pt]
             ~~~~Decay modes ~~ &~~ Branching fractions ($\times 10^{-4}$) ~~~~\\ 
			 \hline
            $X(3872) \phi$  & $1.1 \pm 0.4$  \\
			$X(3872) \eta$  & $0.15 \pm 0.07$  \\
			$X(3872) \eta'$ & $0.17 \pm 0.08$ \\
   \hline
            $\psi(2S) \phi$   & $5.2 \pm 0.4$  \\
			$\psi(2S) \eta$   & $3.3 \pm 0.9$  \\
			$\psi(2S) \eta'$  & $1.29 \pm 0.35$ \\
   \hline
            $J/\psi \phi$   & $10.4 \pm 0.4$ \\
			$J/\psi \eta$   & $4.0 \pm 0.7$ \\
			$J/\psi \eta'$  & $3.3 \pm 0.4$ \\
	\toprule[1.5pt]
		\end{tabular}    
\label{table:Br}
\end{table}

Table~\ref{table:ratio2} shows that the results for $R_2$ are close to one since both $\psi(2S)$ and $J/\psi$ are charmonium states. Furthermore, the obtained ratios of $R_1$ are much smaller than that of $R_2$, which indicates that the $X(3872)$ state is not pure charmonium.

\begin{figure*}[htbp]
	\centering
	\subfigure[]{\includegraphics[scale=0.32]{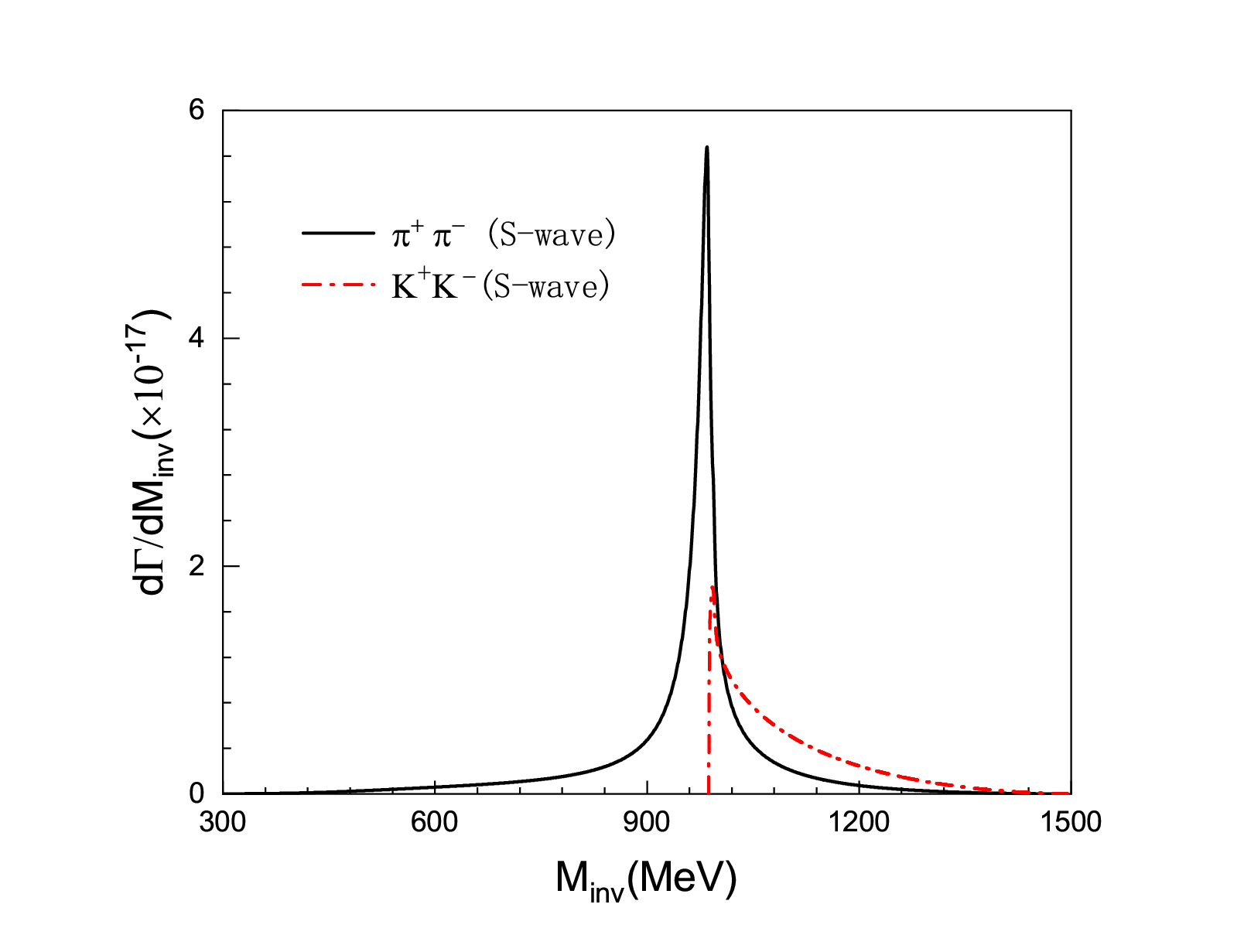}}
	\subfigure[]{\includegraphics[scale=0.32]{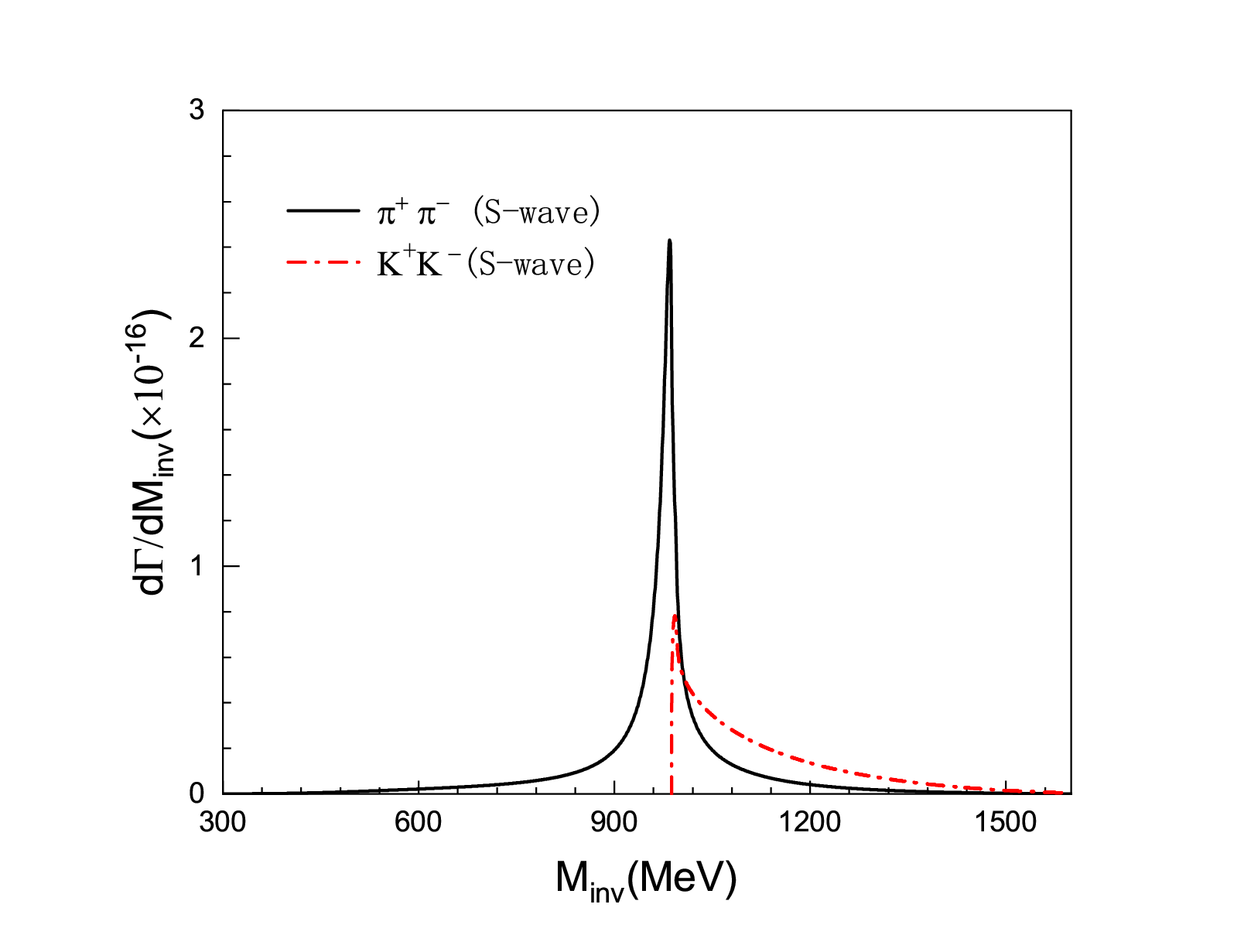}}
	\caption{The $\pi^+\pi^-$ and $K^+K^-$ invariant mass distribution of $B_s^0$ decay with the final state (a) $X(3872)$ and (b) $\psi(2S)$.} 
	\label{fig:pipiKK}
\end{figure*}

Finally, we consider the $B^0_s \to X(3872)[\psi(2S)] K^+K^-$ process. The theoretical results for the invariant $K^+ K^-$ mass distributions are shown in Fig.~\ref{fig:pipiKK}, where the numerical results for the invariant $\pi^+ \pi^-$ mass distributions of the $B^0_s \to X(3872)[\psi(2S)] \pi^+ \pi^-$ decays are also shown. The production rate of the $X(3872)$ is almost an order of magnitude smaller than that of the production of $\psi(2S)$ for both $\pi^+\pi^-$ and $K^+K^-$ final states. The final-state interactions of $\pi^+ \pi^-$ and $K^+K^-$ occur in $s$-wave, where only the $f_0(980)$ contribution is considered. It is expected that future experimental measurements can test these calculations.

It is interesting to compare the branching fractions through the integral of invariant mass $M_{\pi\pi}$ and $M_{KK}$. The results are given by
\be 
\!\!\!\!\!\!\!\! \frac{\mathcal{B}[B_s^0 \to X(3872)(f_0(980) \to K^+K^-)]}{\mathcal{B}[B_s^0 \to X(3872) (f_0(980) \to \pi^+\pi^-)]} &=& 0.5 \pm 0.3, \\
\!\!\!\!\!\!\! \frac{\mathcal{B}[B_s^0 \to \psi(2S) (f_0(980) \to K^+K^-)]}{\mathcal{B}[B_s^0 \to \psi(2S) (f_0(980) \to \pi^+\pi^-)]} &=& 0.6 \pm 0.3,
\label{eq:branching ratio}
\ee 
which shows that the branching fraction obtained from the integral over invariant mass $M_{KK}$ is of the same order of magnitude as that for $M_{\pi\pi}$ while the strength of $K^+K^-$ invariant mass distribution below the peak of $f_0(980)$ is much smaller than that for $\pi^+\pi^-$.

Moreover, it is easy to get the branching fraction from the measurements of Ref.~\cite{LHCb:2020coc},
\be 
    \!\!\!\! \mathcal{B}[B_s^0 \to X(3872) (K^+ K^-)_{\text{non}-\phi}] 
    = (8.6\pm3.5)\times10^{-5}.
\label{eq:non-phi}
\ee 
Then, one can also get the following ratio,
\be 
    \!\!\! \frac{\mathcal{B}[B_s^0 \to X(3872) (f_0(980) \to K^+K^-)]}{\mathcal{B}[B_s^0 \to X(3872)(K^+K^-)_{{\rm non-}\phi}]} 
    = 0.06 \pm 0.02,
\label{eq:KK ratio}
\ee 
which means that the $s$-wave $K^+K^-$ contribution from $f_0(980)$ is extremely small compared with other non-$\phi$ contributions.

For the contribution of the $\phi$ meson to the $B_s^0 \to X(3872) K^+K^-$, with the above obtained couplings of $g_{BX\phi}^2 $ and $g_{\phi KK}^2$, we get the branching fraction
\be 
    \!\!\!\! \mathcal{B}[B_s^0 \to X(3872) (\phi \to K^+K^-)] 
    = (8.3 \pm 3.0)\times 10^{-5},
\label{eq:phi ratio}
\ee 
which is consistent with the following result from the narrow width approximation within the uncertainty,
\be 
&& \mathcal{B}[B_s^0 \to X(3872) (\phi \to K^+K^-)] = \mathcal{B}[B_s^0 \to X(3872)\phi]  \notag \\
    &&  \times \mathcal{B}[\phi \to K^+K^-]  = (5.4 \pm 2.0)\times 10^{-5},
\label{eq:phi ratio2}
\ee 
where we have used $\mathcal{B}[\phi \to K^+K^-] = (49.1 \pm 0.5)\%$ from the RPP~\cite{Workman:2022ynf}.


\section{Summary} \label{sec:summary}

We have investigated the decays of $B_s^0$ into $\psi(2S)\pi^+\pi^-$ and $X(3872)\pi^+\pi^-$ and performed a $\chi^2$-fit to the $\pi^+\pi^-$ invariant mass distributions based on the experimental data from the LHCb collaboration. Taking the dominant Cabibbo favored weak decay mechanism of $B_s^0$, we firstly get $\psi(2S)$ or $X(3872)$ and an $s\bar{s}$ pair. Second, after the hadronization of $s\bar{s}$, we get $\pi^+\pi^-$ and $K^+K^-$ in the final state, and this interaction is mediated by the scalar meson $f_0(980)$. In addition, the contribution from the scalar meson $f_0(1500)$ is also considered for the $B_s^0 \to \psi(2S)\pi^+\pi^-$ decay. It is found that the recent LHCb experimental measurements on the $\pi^+\pi^-$ invariant mass distributions of $B_s^0 \to \psi(2S)\pi^+\pi^-$ decay can be well reproduced. 

Within the same picture, we also studied the $B_s^0 \to X(3872) \pi^+\pi^-$ decay. We find that, to reproduce the experimental data, one needs a different production coupling parameter for $X(3872)$, which indicates that the production of $X(3872)$ is not the same as the production of the charmonium state $\psi(2S)$. Moreover, we have compared the modulus squared of amplitudes of $B_s^0$ decays into $X(3872)$ or $\psi(2S)$ and one light meson, namely $f_0(980)$, $\phi$, $\eta$, and $\eta'$. The results indicate that the production amplitudes of $X(3872)$ in $B_s^0$ decays are different from that of one charmonium in the same $B_s^0$ decays. This may indicate that the $X(3872)$ is not a pure charmonium state.

The $\pi^+ \pi^-$ and $K^+ K^-$ invariant mass distributions for the processes $B_s^0 \to \psi(2S) [X(3872)]\pi^+\pi^-$ and $B_s^0 \to \psi(2S) [X(3872)] K^+ K^-$ are calculated, where we have naturally considered the $K^+ K^-$ final state from $f_0(980)$ for the decays of $B_s^0$ into $\psi(2S)K^+K^-$ and $X(3872)K^+K^-$  in the coupled channel approach, and compared with the $\pi^+\pi^-$ final state in the same situation. On the one hand, it is found that the peak strength of $f_0(980)$ in $m_{\pi\pi}$ is higher than that in $m_{KK}$ for the production of $X(3872)$ or $\psi(2S)$ in the $B_s^0$ decays. On the other hand, we realize that $\mathcal{B}[B_s^0 \to X(3872)\pi^+\pi^-]$ is bigger than $\mathcal{B}[B_s^0 \to X(3872)K^+K^-]$ while both are of the same order of magnitude. The above result does not change with the substitution of $\psi(2S)$ for $X(3872)$. The results here shed light on the fact that the low-lying scalar meson $f_0(980)$ is formed from the interaction of pseudoscalar meson and pseudoscalar meson and that $X(3872)$ is indeed not a pure charmonium state.

\section*{ACKNOWLEDGEMENT}

This work is partly supported by the National Key R\&D Program of China under Grant No. 2023YFA1606703 and by the National Natural Science Foundation of China under Grant Nos. 12075288, 12075133, and 12361141819, and by the Natural Science Foundation of Shandong Province under Grant Nos. ZR2021MA082, and ZR2022ZD26. It is also supported by the Youth Innovation Promotion Association CAS and the Taishan Scholar Project of Shandong Province under Grant No. tsqn202103062.



\end{document}